\documentclass[twocolumn,showpacs,preprintnumbers,amsmath,amssymb]{revtex4}
\usepackage{epsfig}
\usepackage{dcolumn}% Align table columns on decimal point
\usepackage{bm}% bold math

\begin{document}

\title{Sensitivity to new physics: $a_e$ vs. $a_\mu$}

\author{M. V. Chizhov}\thanks{On leave of absence from
Centre for Space Research and Technologies, Faculty of Physics,
University of Sofia, 1164 Sofia, Bulgaria} \affiliation{Departamento
de F\'isica, Universidade Federal da Para\'iba, Caixa Postal 5008,
Jo\~ao Pessoa, PB 58051-970, Brazil.}


\begin{abstract}
At present it is generally believed that ``new physics'' effects
contribute to leptonic anomalous magnetic moment, $a_\ell$, via
quantum loops only and they are proportional to the squared lepton
mass, $m^2_\ell$. An alternative mechanism for a contribution by new
physics is proposed. It occurs at {\em the tree level} and exhibits
{\em a linear} rather than quadratic dependence on $m_\ell$. This
leads to a much larger sensitivity of $a_e$ to the new physics than
was expected so far.
\end{abstract}


\maketitle

\section{Introduction}
Since Schwinger's one-loop calculation~\cite{Schwinger} leptonic
anomalous magnetic moments have usually been used for precision
tests of the Standard Model (SM). Very precise recent experimental
measurements of the electron anomalous magnetic moment~\cite{ae}
\begin{equation}\label{expae}
    a_e^{\rm exp}=1\,159\,652\,180.73(0.28)\times 10^{-12}~[0.24~{\rm ppb}]
\end{equation}
and the muon anomalous magnetic moment~\cite{amu}
\begin{equation}\label{expamu}
    a_\mu^{\rm exp}=1\,165\,920.80(0.63)\times 10^{-9}~[0.54~{\rm ppm}]
\end{equation}
give a possibility to look further for allusive ``new physics''.

Indeed, $a_e^{\rm exp}$ is the most precise experimental value,
which provides a determination of $\alpha$, the fine structure
constant~\cite{alpha}:
\begin{equation}\label{alpha}
    \alpha^{-1}\left(a^{\rm
    exp}_e\right)=137.035\,999\,084(051)~[0.37~{\rm ppb}],
\end{equation}
with an accuracy of more than an order of magnitude better than the
independent measurements~\cite{Rb,Cs}
\begin{eqnarray}\label{valueRb}
    \alpha^{-1}\left(Rb\right)&=&137.035\,998\,78(091)~[6.7~{\rm ppb}],\\
    \label{valueCs}
    \alpha^{-1}\left(Cs\right)&=&137.036\,000\,00(110)~[7.7~{\rm ppb}].
\end{eqnarray}
It is this fact that limits at present testing the $a_e^{\rm SM}$
prediction.

On the other hand the $a_\mu^{\rm exp}$ persists to show a deviation
in comparison with the SM prediction~\cite{eebased}. To be more
definitive we choose a little bit conservative, but the most
recently updated value~\cite{difa}
\begin{equation}\label{Da}
    \Delta a_\mu=a_\mu^{\rm exp}-a_\mu^{\rm th}=+267(96)\times
    10^{-11},
\end{equation}
which shows 2.8~$\sigma$ standard deviation.

Remarkably, this difference exceeds by an order of magnitude the
biggest uncertainties from the hadronic contributions to the muon
anomalous magnetic moment and it is two times larger than the SM
electroweak contribution. The latter fact is apparently in some
conflict with the viable at present ``natural'' conception that new
physics contributions are induced by quantum loop effects, rather
than at the tree level~\cite{Czarnecki}. Thanks to the mass limits
set by LEP and Tevatron, it is highly non-trivial to reconcile the
observed deviation with many of the new physics scenarios. Only the
$\tan\beta$ enhanced contributions in SUSY extensions of the SM for
$\mu > 0$ and/or large enough $\tan\beta$ may explain the ``missing
contribution''.

Based on this approach it is generally expected, that contributions
to the leptonic anomalous magnetic moment are proportional to
$m^2_\ell/\Lambda^2$, where $\Lambda$ is the scale of the new
physics. It leads to the conclusion, that $a_\mu$ is more sensitive
to new physics. The $m^2_\mu/m^2_e\simeq 43\,000$ relative
enhancement for the muon more than compensates for the factor of
$\delta a_\mu^{\rm exp}/\delta a_e^{\rm exp}\simeq 2\,250$ current
experimental precision advantage of $a_e$.

In this paper we consider a model, which allows to generate a
contribution of the new physics to the leptonic anomalous magnetic
moment at {\em the tree level}. Moreover, the contribution exhibits
{\em a linear} rather than quadratic dependence on $m_\ell$. It
changes drastically the situation with the relative sensitivity to
new physics of the muon versus the electron anomalous magnetic
moment. The mass ratio $m_\mu/m_e\simeq 200$ cannot anymore
compensate the advantage of $\delta a_e^{\rm exp}$ over $\delta
a_\mu^{\rm exp}$, which results in a much larger sensitivity of
$a_e$ to the new physics than was expected so far.

\section{The model}
In this paper we are going to investigate the physical consequences
of interacting spin-1 massive bosons described by a formalism of the
second rank antisymmetric tensor fields. The corresponding
Lagrangian, which has been successfully used already during more
than two decades in the chiral perturbation theory, has the
form~\cite{Gasser}
\begin{equation}\label{Vmn}
    {\cal L}_0^T=-\frac{1}{2}\,\partial^\mu T_{\mu\nu}\cdot
    \partial_\rho T^{\rho\nu}
    +\frac{1}{4}\, M^2\, T_{\mu\nu}\, T^{\mu\nu}.
\end{equation}
Using the canonical formalism it can be shown~\cite{Ecker} that the
Lagrangian describes the evolution of the three physical degrees of
freedom of the vector $\left(T_{01},T_{02},T_{03}\right)$, while the
three unphysical components of the axial-vector
$\left(T_{23},T_{31},T_{12}\right)$ do not propagate and they are
frozen.

Although on the mass shell such description of the spin-1 massive
bosons is equivalent to the usual formalism, using vector Proca
fields $V_\mu$, off-shell they have different unphysical states and
can, in general, lead to different physical effects. For example,
the gauge-like Yukawa coupling of the vector field to the bilinear
vector combination of the fermion fields
\begin{equation}\label{YV}
    {\cal L}^V_{\rm int}=g_V\,\bar{\psi}\gamma^\mu\psi\cdot V_\mu
\end{equation}
leads to the well-known static Coulomb interaction due to the exchange
of the unphysical degree of freedom $V_0$. Therefore, the
antisymmetric tensor field, possessing a richer structure of the
unphysical states than the vector field, can give birth to new
physical effects due to its coupling to a corresponding
fermion current.

A simple generalization of the Yukawa coupling (\ref{YV}) in the
case of the antisymmetric tensor field reads
\begin{equation}\label{YT}
    {\cal L}^T_{\rm int}=g_T\,\bar{\psi}\sigma^{\mu\nu}\psi\cdot
    T_{\mu\nu},
\end{equation}
where $\sigma^{\mu\nu}=\frac{i}{2}
\left(\gamma^\mu\gamma^\nu-\gamma^\nu\gamma^\mu\right)$ is the
antisymmetric hermitian matrix. It is interesting to note, that
despite intensive utilization of the original Yukawa interactions
for describing the Higgs boson couplings or the gauge interactions
(\ref{YV}), the interaction (\ref{YT}) still does not have broad
phenomenological applications. Here we would like to discuss one of
its consequences.

\begin{figure}[h]
\begin{picture}(200,70)(-45,20)
% Fig 1(a)
\put(52,50){\circle{40}} \multiput(32,50)(40,0){2}{\circle*{2}}
\put(52,70){\vector(1,0){1}} \put(52,30){\vector(-1,0){1}}
\multiput(74,50)(8,0){4}{\oval(4,5)[t]}
\multiput(78,50)(8,0){4}{\oval(4,5)[b]}
\multiput(0,49)(0,2){2}{\line(1,0){32}} \put(2,57) {\large
$T_{\mu\nu}$} \put(89,57) {\large $V_\alpha$}
\multiput(8,43)(75,0){2}{\vector(1,0){10}}
\multiput(10,35)(75,0){2}{\large $q$}
\end{picture}
\caption{\label{rad}Mixing between the antisymmetric tensor field
and the vector field.}
\end{figure}
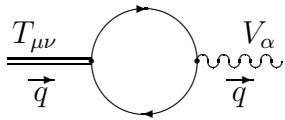

Since the quantum numbers of the physical degrees of freedom of the
vector field $V_i$ (here Latin indices run over $i=1,2,3$) and the
antisymmetric tensor field $T_{0i}$ are the same, they can mix.
Indeed, the quantum loop corrections (see Fig.~\ref{rad})
generate the following additional mixing term
\begin{equation}\label{mix}
    {\cal L}_{\rm int}^{VT}=-\frac{1}{2}\,m_\chi
    \left(\partial^\mu V^\nu-\partial^\nu V^\mu\right)\cdot
    T_{\mu\nu}
\end{equation}
to the total Lagrangian of the interacting vector and antisymmetric
tensor fields. Here
\begin{equation}\label{mchi}
    m_\chi=-i\sum_f\int\frac{{\rm
    d}^4p}{(2\pi)^4}\frac{8g_V^f g_T^f m_f}{(p^2-m^2_f)
    [(p-q)^2-m^2_f]}
\end{equation}
is the effective mass parameter, which leads to the nontrivial
mixing between the antisymmetric tensor field and the vector field
in the case of the chiral symmetry breaking. The summation in
(\ref{mchi}) is performed by all fermion flavors $f$, which couple
simultaneously to the tensor antisymmetric field and to the vector
field, and have also nonzero mass terms $m_f\ne 0$.

An important property of such mixing consists in the gauge invariant
form of the coupling (\ref{mix}) for the vector field $V_\mu$. This
allows to preserve the gauge invariance of the free Lagrangian
\begin{equation}\label{L0V}
    {\cal L}_0^V=-\frac{1}{4}\,F_{\mu\nu}F^{\mu\nu}
\end{equation}
and the zero mass term for the vector field, where as usual
$F_{\mu\nu}=\partial_\mu V_\nu-\partial_\nu V_\mu$ is the gauge
invariant field strength tensor. The resulting mixing between the
antisymmetric tensor field and the vector field is dynamical one,
since it depends on the momentum transfer $q_\mu$. In general, it
leads to very complicated expressions for the physical states after
diagonalization.

In our case it is simplified by the physical conditions of very
small momentum transfers, which we are going to discuss. The second
simplification comes from an assumption of a smallness of the mixing
parameter $m_\chi$ in comparison with very heavy boson mass $M$, so
that their ratio is negligibly small. In this case the only
dominating term in the Lagrangian, including contributions from
(\ref{Vmn}), (\ref{mix}) and (\ref{L0V}), is the mass term from
(\ref{Vmn}) and the procedure of diagonalization consists in a
simple rearrangement of the terms
\begin{eqnarray}\label{diag}
    {\cal L}_0\hspace{-0.15cm}&=&\hspace{-0.15cm}\frac{1}{4}\,M^2
    \!\left(
    T_{\mu\nu}-\frac{m_\chi}{M^2}F_{\mu\nu}\right)
    \hspace{-0.1cm}\left(
    T^{\mu\nu}-\frac{m_\chi}{M^2}F^{\mu\nu}\right)
    \nonumber\\
    &-&\hspace{-0.15cm}\frac{1}{4}\,F_{\mu\nu}F^{\mu\nu}
    \left(1+\frac{m^2_\chi}{M^2}\right).
\end{eqnarray}

Therefore, the physical vector field
\begin{equation}\label{Vph}
    V'_\mu=V_\mu\sqrt{1+\frac{m^2_\chi}{M^2}}
\end{equation}
is defined up to the normalization factor. However, such
transformation does not lead to a physically observable effect,
since it reduces effectively to a redefinition of the coupling
constant $g_V$. On the other side, the physical antisymmetric tensor
field
\begin{equation}\label{Tph}
    T'_{\mu\nu}=T_{\mu\nu}-\frac{m_\chi}{M^2}F_{\mu\nu}
\end{equation}
is defined by the inhomogeneous transformation, which results in the
appearance of the anomalous coupling from the interaction (\ref{YT})
\begin{equation}\label{anom}
    {\cal L}_{\rm int}^{\rm anom}=g_T\frac{m_\chi}{M^2}
    \,\bar{\psi}\sigma^{\mu\nu}\psi\cdot F_{\mu\nu}
\end{equation}
and the corresponding anomalous magnetic moment for the fermion
field
\begin{equation}\label{aF}
    a_\psi=4\,\frac{g_T}{g_V}\,\frac{m_\chi}{M^2}\,m_\psi.
\end{equation}

\section{The experimental consequences}

In the previous section we have shown, that an additional
contribution from the new physics to an anomalous magnetic moment of
the fermion can be generated at the tree level. The role of a new
physics here is played by the non-trivial coupling (\ref{YT}) of the
massive spin-1 boson, described by the antisymmetric tensor field,
to the fermion tensor current. This coupling leads inevitably to the
mixing (\ref{mix}) between the known gauge fields, such as the
photon, and the new hypothetical spin-1 heavy boson. The smallness
of the mixing parameter $m_\chi$ and the heaviness of the new boson
mass $M$ could be the reasons why their effects and the direct
production of such particles have not been registered up to now.

Probably the only places where such effect could be tested in
low-energy physics are the very precise measurements of the
anomalous photon couplings to the leptons, namely electron and muon.
Therefore, the difference (\ref{Da}) between the predicted and the
measured anomalous magnetic moment of the muon may be explained {\em
completely\/} by the new mechanism, if the following identification
holds
\begin{equation}\label{Damu}
    \Delta a_\mu=4\,\frac{g^\mu_T}{e}\,\frac{m_\chi}{M^2}\,m_\mu.
\end{equation}

Unfortunately, the only one experimentally measured value cannot fix
separately each of the three new parameters $g^\mu_T$, $m_\chi$ and
$M$. Nevertheless, our predictions can be more definitive, if we
make an additional assumption about the universality of the new
Yukawa coupling constant $g_T$. Let us assume, that by an analogy
with the gauge coupling $g_V$, which is the same for different
fermion generations, the new coupling $g_T$ also possesses the
universality condition
\begin{equation}\label{leptuniv}
    g_T=g^e_T=g^\mu_T=g^\tau_T.
\end{equation}

In this case the contribution of the new physics to the anomalous
magnetic moment of the lepton
\begin{equation}\label{al}
    \Delta a_\ell=\kappa\,m_\ell
\end{equation}
depends linearly on the lepton mass, where the coefficient
\begin{equation}\label{const}
    \kappa=4\,\frac{g_T}{e}\,\frac{m_\chi}{M^2}=(25.3\pm 9.1)\times
    10^{-12}{~\rm MeV}^{-1}
\end{equation}
is assumed to be universal for  each lepton species.

Therefore, we are in position now to make a definitive prediction
for a new physics effect on the electron anomalous magnetic moment
$a_e$. The linear (\ref{al}) rather than quadratic dependence on
$m_\ell$ results in a huge effect due to  the new physics on the
determination of the fine structure constant $\alpha$ via $a_e$. So,
according to the formula (\ref{al}) there should be an  additional
contribution
\begin{equation}\label{Dae}
    \Delta a_e=(12.9\pm 4.6)\times 10^{-12}
\end{equation}
to the anomalous magnetic moment of the electron from the new
physics, which is well above the non-QED contributions $a^{\rm
HAD}_e = 1.671(19)\times 10^{-12}$, $a^{\rm EW}_e = 0.030(01)\times
10^{-12}$ \cite{Mohr} and the experimental precision $\delta a^{\rm
exp}_e = 0.28\times 10^{-12}$ \cite{ae}.

\begin{figure}[ht]
\epsfig{file=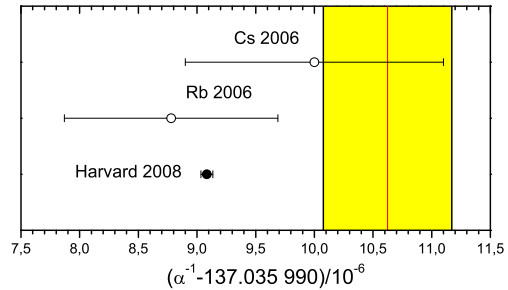,width=8cm} \caption{\label{fig:2} The most
accurate $\alpha$ determination and our prediction.}
\end{figure}

If we subtract the additional contribution (\ref{Dae}) from the
experimentally measured value $a^{\rm exp}_e$ (\ref{expae}), this
results in a lower value of the fine structure constant than the
extracted one (\ref{alpha}). Indeed we predict, that the inverse
value of $\alpha$ should be on
\begin{equation}\label{Dalpha}
    \Delta \alpha^{-1}=(1.52\pm 0.55)\times 10^{-6}
\end{equation}
greater than presently accepted (Fig.~\ref{fig:2}).
Unfortunately, our prediction cannot be verified at present, because
the independent $\alpha$ determinations
(\ref{valueRb},\ref{valueCs}) have uncertainties comparable with the
contribution (\ref{Dalpha}).

Beside of the description of the absolute value of the difference
between the predicted and measured anomalous magnetic moment of the
muon, it is interesting also to predict its sign. It could be done
in our framework, if we make further assumptions. Let us assume,
that the new massive boson interacts only with the {\em down\/}-type
fermions and, by an analogy with the electric charge, all coupling
constants $g^{\rm down}_T$ have the same sign. In this case the
generated coefficient (\ref{mchi}) in the mixing term multiplied by
the ratio $g_T/e$ results in the positive constant $\kappa$.
Therefore, it confirms that the experimental value for the muon
anomalous magnetic moment is higher than the predicted one. It is
interesting also to note that if the new boson exists and it is not
too heavy, $M < 3$~TeV, it may be observed in the Drell--Yan process
at the LHC.

\section{Conclusions}

In this paper we have considered the alternative scenario for a
contribution by the new physics to the leptonic anomalous magnetic
moment. The key role in this scenario belongs to a new massive
spin-1 boson, which is described by a second rank antisymmetric
tensor field. The latter has new non-minimal tensor interactions
with fermions that lead to its mixing with the photon in the case of
a chirally broken symmetry. Therefore, the initial wave functions of
the antisymmetric tensor field and the photon can be expressed
through linear combinations of their physical states, which results
in the appearance of  a direct anomalous photon coupling to the
fermions at the tree level.

In the case of universality of the new tensor interactions the
contribution of the new physics to the anomalous magnetic moment of
the lepton depends linearly on the lepton mass. This leads to a
higher sensitivity of the electron anomalous magnetic moment to the
new physics than was expected before. The latter fact may
substantially affect the extraction of a real value of the fine
structure constant from $a_e$.

\section*{Acknowledgements}
We are grateful to S. Eidelman and Z. Zhang for the fruitful
correspondence.

This work was financially supported by the Brazilian agency CAPES.

%\pagebreak[3]

\end{document}